\documentclass[smallextended, envcountsect]{svjour3}       % onecolumn (second format)

\smartqed  % flush right qed marks, e.g. at end of proof

\usepackage{amsfonts}                               % Real (R) symbol
\usepackage{bm}                                     % bold in equations
\usepackage{booktabs}                               % professional-quality tables
\usepackage[nolist]{acronym}                        % acronyms
\usepackage{graphicx}
\usepackage[utf8]{inputenc}                         % allow utf-8 input
\usepackage{lmodern}                                % https://tex.stackexchange.com/a/58088/216202
\usepackage{multirow}                               % for tables
\usepackage{nicefrac}                               % compact symbols for 1/2, etc.
\usepackage{siunitx}                                % large numbers, SI numbers
\usepackage{subcaption}
\usepackage{xcolor}                                 % colored fonts
\usepackage{amsmath}

\usepackage{hyperref}
\usepackage[capitalise]{cleveref}

\newcommand{\myhref}[1]{\href{#1}{\texttt{#1}}}

\newcommand{\comment}[1]{}

\DeclareMathOperator*{\argmin}{arg\,min}

\journalname{IJCARS}

\hypersetup{final}

\graphicspath{{img/}}

\makeatletter
\newcommand*{\org@overidelabel}{}
\let\org@overridelabel\@verridelabel
\@ifpackagelater{acronym}{2015/03/21}{% v1.41
  \renewcommand*{\@verridelabel}[1]{%
    \@bsphack
    \protected@write\@auxout{}{\string\AC@undonewlabel{#1@cref}}%
    \org@overridelabel{#1}%
    \@esphack
  }%
}{% older versions
  \renewcommand*{\@verridelabel}[1]{%
    \@bsphack
    \protected@write\@auxout{}{\string\undonewlabel{#1@cref}}%
    \org@overridelabel{#1}%
    \@esphack
  }%
}
\makeatother

\definecolor{rev1}{HTML}{000000}  % for clean version

\begin{document}
    \begin{acronym}
    \acro{ad}[AD]{Alzheimer's Disease}
    \acro{adni}[ADNI]{\ac{ad} Neuroimaging Initiative}
    \acro{amp}[AMP]{automatic mixed precision}
    \acro{bite}[BITE]{Brain Images of Tumors for Evaluation}
    \acro{cnn}[CNN]{convolutional neural network}
    \acro{dsc}[DSC]{Dice score coefficient}
    \acro{ez}[EZ]{epileptogenic zone}
    \acro{gbm}[GMB]{glioblastoma}
    \acro{ixi}[IXI]{Information eXtraction from Images}
    \acro{mni}[MNI]{Montreal Neurological Institute}
    \acro{nhnn}[NHNN]{National Hospital for Neurology and Neurosurgery}
    \acro{nhs}[NHS]{National Health Service}
    \acro{oasis}[OASIS]{Open Access Series of Imaging Studies}
    \acro{prelu}[PReLU]{Parametric Rectified Linear Unit}

    \acro{csf}[CSF]{cerebrospinal fluid}
    \acro{ct}[CT]{computerized tomography}
    \acro{gif}[GIF]{geodesical information flows}
    \acro{mr}[MR]{magnetic resonance}
    \acro{mri}[MRI]{magnetic resonance image}
    \acro{cqv}[CQV]{coefficient of quartile variation}
    \acro{tta}[TTA]{test-time augmentation}
    \acro{ttd}[TTD]{test-time dropout}
    \acro{t1}[$T_1$w]{$T_1$-weighted}
    \acro{t2}[$T_2$w]{$T_2$-weighted}
    \acro{t1gad}[$T_1$wCE]{\ac{t1} \ac{mri} with gadolinium contrast enhancement}
    \acro{flair}[FLAIR]{fluid-attenuated inversion recovery}
    \acro{nmi}[NMI]{normalized mutual information}
    \acro{rc}[RC]{resection cavity}
    \acrodefplural{rc}[RCs]{resection cavities}
    \acro{rs}[RS]{random simulation}
    \acro{tpo}[TPO]{temporal-parietal-occipital}

    \acro{pip}[PIP]{Pip Installs Packages}
    \acro{pypi}[PyPI]{Python Package Index}
\end{acronym}

    \title{%
    A self-supervised learning strategy
    for postoperative brain cavity segmentation
    simulating resections
}

\titlerunning{%
    A self-supervised learning strategy for cavity segmentation simulating resections
} % if too long for running head

\author{
    Fernando Pérez-García\textsuperscript{1,2,3} \and
    Reuben Dorent\textsuperscript{3} \and
    Michele Rizzi\textsuperscript{4} \and
    Francesco Cardinale\textsuperscript{4} \and
    Valerio Frazzini\textsuperscript{5,6,7} \and
    Vincent Navarro\textsuperscript{5,6,7} \and
    Caroline Essert\textsuperscript{8} \and
    Irène Ollivier\textsuperscript{9} \and
    Tom Vercauteren\textsuperscript{3} \and
    Rachel Sparks\textsuperscript{3} \and
    John S. Duncan\textsuperscript{10,11} \and
    Sébastien Ourselin\textsuperscript{3}
}

\authorrunning{Fernando Pérez-García et al.}

\institute{
    Fernando Pérez-García -- \email{fernando.perezgarcia.17@ucl.ac.uk} \\
    \textsuperscript{1}Department of Medical Physics and Biomedical Engineering, UCL, London, UK \\
    \textsuperscript{2}Wellcome / EPSRC Centre for Interventional and Surgical Sciences, UCL, London, UK \\
    \textsuperscript{3}School of Biomedical Engineering \& Imaging Sciences, King's College London, London, UK \\
    \textsuperscript{4}``C. Munari'' Epilepsy Surgery Centre ASST GOM Niguarda, Milan, Italy \\
    \textsuperscript{5}Paris Brain Institute, ICM, INSERM, CNRS, F-75013, Paris, France \\
    \textsuperscript{6}Sorbonne Université, F-75013, Paris, France \\
    \textsuperscript{7}AP-HP, Pitié-Salpêtrière Hospital, Epilepsy Unit, Reference Center for Rare Epilepsies, and Departement of Clinical Neurophysiology, F-75013, Paris, France \\
    \textsuperscript{8}Université de Strasbourg, CNRS, ICube, Strasbourg, France \\
    \textsuperscript{9}Department of Neurosurgery, Strasbourg University Hospital, Strasbourg, France \\
    \textsuperscript{10}UCL Queen Square Institute of Neurology, London, UK \\
    \textsuperscript{11}National Hospital for Neurology and Neurosurgery, London, UK \\
}

\date{Received: 1 February 2021 / Accepted: 21 May 2021}

\maketitle

    \begin{abstract}
    \textit{Purpose}
    Accurate segmentation of brain \acp{rc} aids in postoperative analysis and determining follow-up treatment.
    \Acp{cnn} are the state-of-the-art image segmentation technique, but require large annotated datasets for training.
    Annotation of 3D medical images is time-consuming, requires highly-trained raters, and may suffer from high inter-rater variability.
    Self-supervised learning strategies can leverage unlabeled data for training.
    \textit{Methods}
    We developed an algorithm to simulate resections from preoperative \acp{mri}.
    We performed self-supervised training of a 3D \ac{cnn} for \ac{rc} segmentation using our simulation method.
    We curated EPISURG, a dataset comprising 430 postoperative and 268 preoperative \acp{mri} from 430 refractory epilepsy patients who underwent resective neurosurgery.
    We fine-tuned our model on three small annotated datasets from different institutions and on the annotated images in EPISURG, comprising 20, 33, 19 and 133 subjects.
    \textit{Results}
    The model trained on data with simulated resections obtained median (interquartile range) \acp{dsc} of 81.7 (16.4), 82.4 (36.4), 74.9 (24.2) and 80.5 (18.7) for each of the four datasets.
    After fine-tuning, \acp{dsc} were 89.2 (13.3), 84.1 (19.8), 80.2 (20.1) and 85.2 (10.8).
    For comparison, inter-rater agreement between human annotators from our previous study was 84.0 (9.9).
    \textit{Conclusion}
    We present a self-supervised learning strategy for 3D \acp{cnn} using simulated \acp{rc} to accurately segment real \acp{rc} on postoperative \ac{mri}.
    Our method generalizes well to data from different institutions, pathologies and modalities.
    Source code, segmentation models and the EPISURG dataset are available at
    \myhref{https://github.com/fepegar/resseg-ijcars}.

\end{abstract}

% Reset acronyms definitions
\acresetall

    \section{Introduction}

\subsection{Motivation}

%% Background

% An accurate quantitative analysis of postoperative images is necessary to understand the relation between presurgical assessment, surgical treatment and patient outcomes.

Approximately one third of epilepsy patients are drug-resistant.
If the \ac{ez}, i.e., ``the area of cortex indispensable for the generation of clinical seizures''~\cite{rosenow_presurgical_2001}, can be localized,
resective surgery to remove the \ac{ez} may be curative.
Currently, 40\% to 70\% of patients with refractory focal epilepsy are seizure-free after surgery~\cite{jobst_resective_2015}.
This is, in part, due to limitations identifying the \ac{ez}.
Retrospective studies relating presurgical clinical features and resected brain structures to surgical outcome provide useful insight to guide \ac{ez} resection~\cite{jobst_resective_2015}. %~\cite{galovic_association_2019}.
To quantify resected structures, first, the resection cavity must be segmented on the postoperative \ac{mri}.
A preoperative image with a corresponding brain parcellation can then be registered to the postoperative \ac{mri} to identify resected structures.

%% Resection of brain tumors
\Ac{rc} segmentation is also necessary in other applications.
For neuro-oncology, the gross tumor volume,
which is the sum of the \ac{rc} and residual tumor volumes,
is estimated for postoperative radiotherapy~\cite{ermis_fully_2020}.

%% Summary of what is missing
Despite recent efforts to segment \acp{rc} in the context of brain cancer~\cite{meier_automatic_2017,ermis_fully_2020}, little research has been published in the context of epilepsy surgery.
Furthermore, previous work is limited by the lack of benchmark datasets, released code or trained models, and evaluation is restricted to single-institution datasets used for both training and testing.

\subsection{Related works}

%% Joint registration and segmentation
After surgery, \acp{rc} fill with \ac{csf}.%~\cite{winterstein_partially_2010}.
This causes an inherent uncertainty in delineating \acp{rc} adjacent to structures such as sulci, ventricles or edemas.
%Moreover, brain shift can occur during surgery, causing either \ac{csf} filling in regions outside of the resection cavity or changes to the brain structures shape and volume.
%Changes in brain position, caused by brain shift, remove the possibility of using symmetry measurements around the sagittal plane or linear registration with the preoperative image to locate the resection cavity.
%These changes impede using symmetry measurements or linear registration with a preoperative image to locate the resection cavity.
Nonlinear registration has been presented to segment the \ac{rc} for epilepsy~\cite{chitphakdithai_non-rigid_2010} and brain tumor~\cite{chen_deformable_2015} surgeries by detecting non-corresponding regions between pre- and postoperative images.
However, evaluation of these methods was
%limited to only six 3D \ac{t1} \acp{mri} from a private dataset and two 2D \ac{mri} slices, respectively.
restricted to a very small number of images.
%Furthermore, in cases with large brain shift or edemas, non-corresponding voxels detected in the image may be either in the resection cavity or in regions with changes after resection.
Furthermore, in cases with intensity changes due to the resection (e.g., brain shift, atrophy, fluid filling), non-corresponding voxels may not correspond to the \ac{rc}.

% %% Deep learning
% \Acp{cnn} have become the state of the art for medical image segmentation~\cite{simpson_large_2019,singh_3d_2020}.
% They have repeatedly shown super-human accuracy in fully-supervised learning settings using massive annotated datasets~\cite{he_delving_2015}.
% However, the performance of neural networks trained
% with fully-supervised learning using small datasets is often poor.
% Annotated medical imaging datasets are often small due to the financial and time burden annotating the (often three-dimensional) data, and the need for highly-trained raters.
% In self-supervised learning, training instances are generated automatically using unlabeled data from a source domain to learn features that can be transferred to a target domain~\cite{jing_self-supervised_2019}.
% Synthetic data can be generated cheaply to perform self-supervised training~\cite{nikolenko_synthetic_2019}.
% A good compromise between supervised and unsupervised learning is training the neural network using a mix of labeled and unlabeled data, referred to as semi-supervised learning~\cite{pan_survey_2010}.
% These techniques can be used to leverage unlabeled medical imaging data to improve training in settings where acquiring annotations is time-consuming or costly.

%% Machine learning for resection cavity
Decision forests were presented for brain cavity segmentation after glioblastoma surgery, using four \ac{mri} modalities~\cite{meier_automatic_2017}.
These methods, which aggregate hand-crafted features extracted from all  modalities to train a classifier, can be sensitive to signal inhomogeneity and unable to distinguish regions with intensity patterns similar to \ac{csf} from \acp{rc}.
Recently, a 2D \ac{cnn} was trained to segment the \ac{rc} on \ac{mri} slices in 30 glioblastoma patients~\cite{ermis_fully_2020}.
%The final 3D segmentation was computed by averaging predictions across anatomical axes.
They obtained a `median (interquartile range)' \ac{dsc} of 84 (10) compared to ground-truth labels by averaging predictions across anatomical axes to compute the 3D segmentation.
While these approaches require four modalities to segment the resection cavity, some of the modalities are often unavailable in clinical settings~\cite{dorent_learning_2021}.
Furthermore, code and datasets are not publicly available, hindering a fair comparison across methods.
Applying these techniques requires curating a dataset with manually obtained annotations to train the models, which is expensive.

%% Self-supervised learning
Unsupervised learning methods can leverage large, unlabeled medical image datasets during training.
In self-supervised learning, training instances are generated automatically from unlabeled data and used to train a model to perform a pretext task. %such as inpainting or image restoration.
The model can be fine-tuned on a smaller labeled dataset to perform a downstream task~\cite{chen_self-supervised_2019}.
The pretext and downstream tasks may be the same.
For example, a \ac{cnn} was trained to reconstruct a skull bone flap by simulating craniectomies on CT scans~\cite{matzkin_self-supervised_2020}.
Lesions simulated in chest CT of healthy subjects were used to train models for nodule detection, improving accuracy compared to training on a smaller dataset of real lesions~\cite{pezeshk_seamless_2017}.

% %% Semi-supervised learning
% Semi-supervised learning may be used when a large amount of unlabeled data is available.
% A model trained on a labeled dataset (which may have been generated in a self-supervised setting) can generate pseudolabels for unlabeled data.
% Uncertainty estimation may be used to select pseudolabeled instances with a low uncertainty for medical image segmentation tasks, improving model performance compared to using a random subset~\cite{venturini_uncertainty_2020}.

\subsection{Contributions}

%% What we present
We present a self-supervised learning approach to train a 3D \ac{cnn} to segment brain \acp{rc} from \ac{t1} \ac{mri} without annotated data, by simulating resections during training.
We ensure our work is reproducible by releasing
%the scripts used for training, the trained \acp{cnn}, the installable Python packages to simulate resections and perform inference, and the dataset used for evaluation.
the source code for resection simulation and \ac{cnn} training, the trained \ac{cnn}, and the evaluation dataset.
To the best of our knowledge, we introduce the first open annotated dataset of postoperative \ac{mri} for epilepsy surgery.

%% Comparison to MICCAI paper
This work extends our conference paper~\cite{perez-garcia_simulation_2020} as follows:
    % \item we used uncertainty estimation as a selection criterion for pseudolabeled instances within our semi-supervised learning setting; and
% introduced uncertainty estimation as selection criteria for semi-supervised learning,
1) we performed a more comprehensive evaluation, assessing the effect of the resection simulation shape on performance and evaluating datasets from different institutions and pathologies;
2) we formalized our transfer learning strategy.

%% Paper organization
%The rest of this paper is organized as follows.
%\Cref{sec:methods} describes the self-supervised training paradigm and our brain resection simulation.
%\cref{sec:experiments_and_results} reports our experimental design and results.
%presents experiments to evaluate our proposed method on simulated and real resection data.
%Finally, \cref{sec:discussion} discusses the results, future directions and potential applications.

\section{Methods}
\label{sec:methods}

\newcommand{\p}{\bm{p}}
\newcommand{\vv}{\bm{v}}
\newcommand{\X}{\bm{X}}
\newcommand{\Y}{\bm{Y}}
\newcommand{\M}{\bm{M}}
\newcommand{\U}{\bm{U}}
\newcommand{\R}{\mathbb{R}}

\newcommand{\img}[2]{#1 : \Omega \to #2}
\newcommand{\binimg}[1]{\img{#1}{ \{ 0, 1 \} }}

\newcommand{\Dom}{\mathcal{D}}
\newcommand{\Tas}{\mathcal{T}}
\newcommand{\Xdo}{\mathcal{X}}
\newcommand{\Ydo}{\mathcal{Y}}
\newcommand{\fp}[1]{f_{\bm{\theta} \text{#1}}}
\newcommand{\wt}{\widetilde}

% https://tex.stackexchange.com/a/466437/216202
\newcommand*\st[1]{_{\textnormal{#1}}}

\newcommand{\post}{\st{post}}
\newcommand{\pre}{\st{pre}}
\newcommand{\cav}{\st{cavity}}
\newcommand{\simul}{\st{sim}}
\newcommand{\lab}{\st{labeled}}
\newcommand{\unl}{\st{unlabeled}}

\subsection{Learning strategy}
\label{sec:learning_strategy}

\subsubsection{Problem statement}

\newcommand{\loss}{\mathcal{L}}
\newcommand{\expec}{\mathbb{E}}
\newcommand{\exppost}{\expec_{\Dom\post}}

The overall objective is to automatically segment \acp{rc} from postoperative \ac{t1} \ac{mri} using a \ac{cnn} $f_{\bm{\theta}}$ parameterized by weights $\bm{\theta}$.
Let $\X_{\text{post}} : \Omega \to \R$ and $\Y\cav  : \Omega \to \{ 0, 1 \}$ be a postoperative \ac{t1} \ac{mri} and its cavity segmentation label, respectively, where $\Omega \subset \R^3$.
$\X_{\text{post}}$ and $\Y_{\text{cavity}}$ are drawn from the data distribution $\Dom\post$.
In model training, the aim is to minimize the expected discrepancy between the label $\Y\cav$ and network prediction $f_{\bm{\theta}}(\X\post)$.
Let $\loss$ be a loss function that estimates this discrepancy (e.g., Dice loss).
The optimization problem for the network parameters $\bm{\theta}$ is:
\begin{equation}
    \bm{\theta}^* =
    \argmin_{\bm{\theta}}
    \exppost \left[
        \loss \left(
            f_{\bm{\theta}} \left( \X\post \right),
            \Y\cav
        \right)
    \right]
    \label{eq:problem_optimization}
\end{equation}

In a fully-supervised setting, a labeled dataset $D\post = \{ (\X_{\text{post}_i}, \Y_{\text{cavity}_i}) \}_{i = 1}^{n\post}$ is employed to estimate the expectation defined in \eqref{eq:problem_optimization} as:
\begin{equation}
    \exppost \left[
        \loss \left(
            f_{\bm{\theta}} \left( \X\post \right), \Y\cav
        \right)
    \right]
    \approx \frac{1}{n\post} \sum_{i=1}^{n\post} \loss(f_{\bm{\theta}}(\X_{\text{post}_i}), \Y_{\text{post}_i})
    \label{eq:problem_optimization_fully}
\end{equation}

In practice, \acp{cnn} typically require an annotated dataset with a large $n\post$ to generalize well for unseen instances.
However, given the time and expertise required to annotate scans, $n\post$ is often small.
We present a method to artificially increase $n\post$ by simulating postoperative \acp{mri} and associated labels from preoperative scans.

\subsubsection{Simulation for domain adaptation and self-supervised learning}
\label{sec:sim_res_self}

Let $D\pre = \{ \X_{\text{pre}_i} \}_{i = 1}^{n\pre}$ be a dataset of preoperative \ac{t1} \ac{mri}, drawn from the data distribution $\Dom\pre$.
We propose to generate a simulated postoperative dataset $D\simul = \{ (\X_{\text{sim}_i}, \Y_{\text{sim}_i}) \}_{i = 1}^{n\simul}$ using the preoperative dataset $D\pre$.
Specifically, we aim to build a generative model $\phi\simul : \X\pre \mapsto (\X\simul, \Y\simul)$ that transforms preoperative images into simulated, annotated postoperative images that imitate instances drawn from the postoperative data distribution $\Dom\post$.
$D\simul$ can then be used to estimate the expectation in \eqref{eq:problem_optimization}:
\begin{equation}
    \exppost \left[\
        \loss\left(
            f_{\bm{\theta}} \left(\X\post \right), \Y\cav \right)
        \right]
        \approx \frac{1}{n\simul}\sum_{i=1}^{n\simul} \loss(f_{\bm{\theta}}(\X_{\text{sim}_i}),  \Y_{\text{sim}_i})
    \label{eq:problem_optimization_sim}
\end{equation}

Simulated images can be generated from any unlabeled preoperative dataset.
Therefore, the size of the simulated dataset can be much greater than the annotated dataset $D\post$, i.e., $n\simul\gg n\post$.
The network parameters $\bm{\theta}$ can be optimized by minimizing \eqref{eq:problem_optimization_sim} using stochastic gradient descent, leading to a trained predictive function $f_{\bm{\theta_}{\text{sim}}}$.
Finally, $f_{\bm{\theta_}{\text{sim}}}$ can be fine-tuned on $D\post$ to improve performance on the postoperative domain $\Dom\post$.

\subsection{Resection simulation for self-supervised learning}
\label{sec:simulation}

\newcommand{\AAA}{\bm{A}}
\newcommand{\NN}{\mathcal{N}}

$\phi\simul$ takes images from $\Dom\pre$ to generate training instances by simulating a realistic shape, location and intensity pattern for the \ac{rc}.
We present simulation of cavity shape and label in \cref{sec:cavity,sec:cavity_constrain}, respectively.
In \cref{sec:texture_cavity}, we present our method to generate the resected image.

\subsubsection{Initial cavity shape}
\label{sec:cavity}

To simulate a realistic \ac{rc}, we consider its topological and geometric properties: it is a single volume with a non-smooth boundary.
We generate a geodesic polyhedron with frequency $f$ by subdividing the edges of an icosahedron $f$ times and projecting each vertex onto a parametric sphere with a unit radius centered at the origin.
This polyhedron models a spherical surface $S = \{ V, F \}$ with vertices
$
    V = \left\{
        \vv_i \in \R^3
    \right\}
    _{i = 1}^{n_V}
$
and faces
$
    F = \left\{
        \bm{f}_k \in \mathbb{N}^3
    \right\}
    _{k = 1}^{n_F}
$, where $n_V$ and $n_F$ are the number of vertices and faces, respectively.
Each face $\bm{f}_k = \{ i_1^k, i_2^k, i_3^k \}$ is a sequence of three non-repeated vertex indices.

To create a non-smooth surface, $S$ is perturbed with simplex noise~\cite{perlin_improving_2002}, a procedural noise generated by interpolating pseudorandom gradients on a multidimensional simplicial grid.
We chose simplex noise as it simulates natural-looking textures or terrains and is computationally efficient for multiple dimensions.
The noise $\eta : \R^3 \to [-1, 1]$ at point $\p \in \R^3$ is a weighted sum of the noise contribution for $\omega$ different octaves, with weights $\{\gamma ^ {n - 1}\}_{n = 1}^{\omega}$ controlled by the persistence parameter $\gamma$.
The displacement $\delta$ of a vertex $\vv$ is:
\begin{equation}
    \delta(\vv)
    = \eta \left( \frac{\vv + \bm{\mu} }{\zeta}, \omega, \gamma \right)
\end{equation}
where
$\zeta$ is a scaling parameter to control smoothness
and $\bm{\mu}$ is a shifting parameter that adds stochasticity
(equivalent to a random number generator seed).
Each vertex $\vv_i$ is displaced radially to create a perturbed sphere:
$
V_{\delta}
    = \left\{
    \vv_i
    + \delta(\vv_i)
    \frac{\vv_i}{\|\vv_i\|}
    \right\}
    _{i = 1}^{n_V}
    = \left\{
    \vv_{\delta i}
    \right\}
    _{i = 1}^{n_V}
$.

Next, a series of transforms is applied to $V_{\delta}$ to modify the mesh's volume and shape.
To add stochasticity, random rotations around each axis are applied to $V_{\delta}$ with the rotation transform
$T\st{R}(\bm{\theta}\st{r}) = R_x(\theta_x) \circ R_y(\theta_y) \circ R_z(\theta_z)$,
where~$\circ$~indicates a transform composition and
$R_i(\theta_i)$ is a rotation of $\theta_i$ radians around axis $i$.
$T\st{S}(\bm{r})$ is a scaling transform,
where $(r_1, r_2, r_3) = \bm{r}$ are semiaxes of an ellipsoid
with volume $v$ used to model the cavity shape.
The semiaxes are computed as
$r_1 = r$, $r_2 = \lambda r$ and $r_3 = r /\lambda$,
where $r = (3 v / 4)^{1/3}$ and
$\lambda$ controls the semiaxes length ratios\footnote{
    Note the volume of an ellipsoid with semiaxes $(a, b, c)$ is $v = \frac{4}{3} \pi a b c$.
}.
These transforms are applied to $V_{\delta}$ to define the initial resection cavity surface $S\st{E} = \{ V\st{E}, F \}$, where
$V\st{E} =
\{
    T\st{S}(\bm{r})
    \circ T\st{R}(\bm{\theta}\st{r})(
        \vv_{\delta i})
\}
_{i = 1}^{n_V}
$.

\subsubsection{Cavity label}
\label{sec:cavity_constrain}

\begin{figure}
    \centering
    % \captionsetup[subfigure]{justification=centering}
    \begin{subfigure}{0.3\textwidth}
        \includegraphics[width=0.8\linewidth]{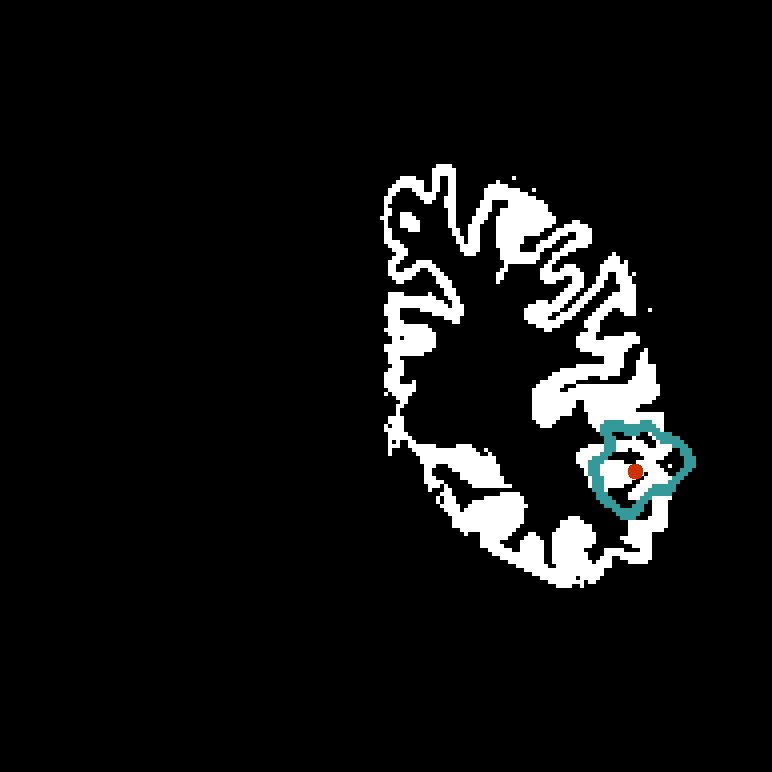}
        \caption{$S_a$ on $\M\st{GM}^h$\label{fig:sama}}
    \end{subfigure}
    \begin{subfigure}{0.3\textwidth}
        \includegraphics[width=0.8\linewidth]{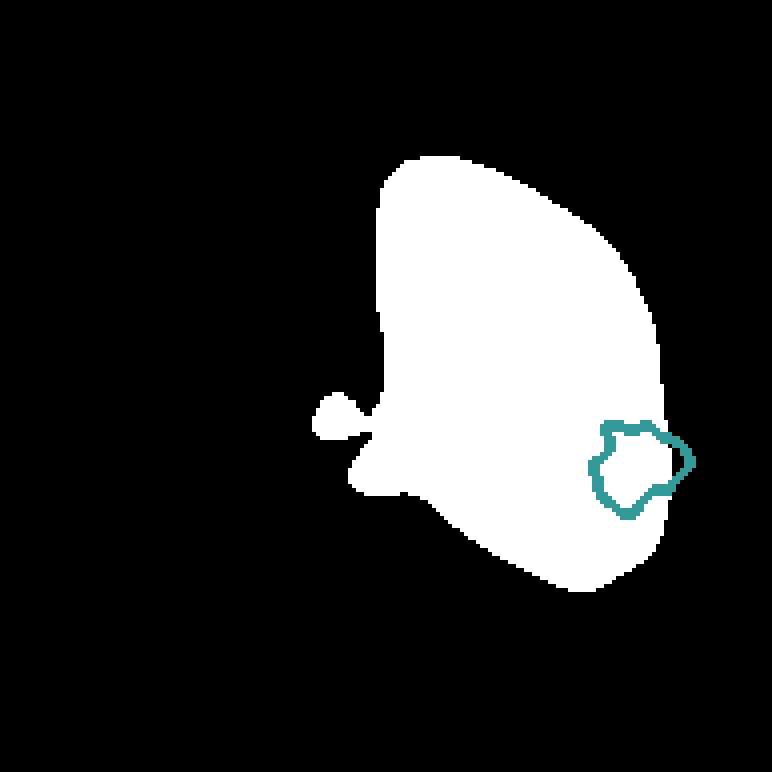}
        \caption{$S_a$ on $\M\st{R}^h$\label{fig:samb}}
    \end{subfigure}
    \begin{subfigure}{0.3\textwidth}
        \includegraphics[width=0.8\linewidth]{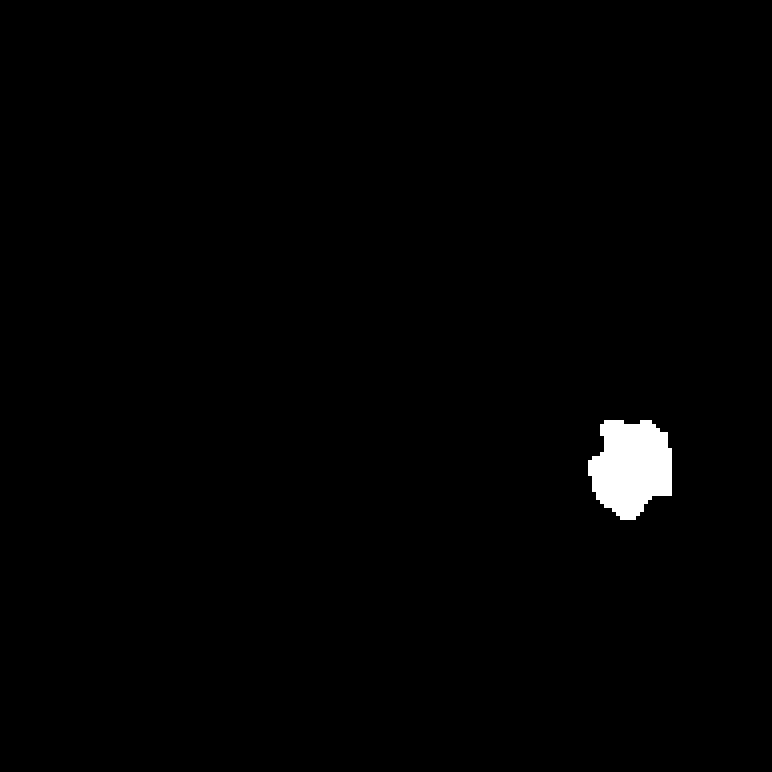}
        \caption{$\Y\simul = \M_{S_a} \odot \M\st{R}^h$\label{fig:mr}}
    \end{subfigure}

    \caption{%
        Simulation of the ground-truth cavity label.
        $S_a$ (blue) is computed by centering $S\st{E}$ on $\bm{a}$, a random positive voxel (red) of $\M\st{GM}^h$ (\subref{fig:sama}).
        $\M_{S_a}$ is a binary mask derived from $S_a$.
        $\Y\simul$ (\subref{fig:mr}) is the intersection of $\M_{S_a}$ and $\M\st{R}^h$ (\subref{fig:samb})
    }
    \label{fig:shape}
\end{figure}

The simulated \ac{rc} should not span both hemispheres or include extracerebral tissues such as bone or scalp.
This section describes our method to ensure that the \ac{rc} appears in anatomically plausible regions.

A \ac{t1} \ac{mri} is defined as $\X\pre : \Omega \to \R$.
A full brain parcellation $\bm{P} : \Omega \to Z$ is generated~\cite{cardoso_geodesic_2015} for $\X\pre$,
where $Z$ is the set of segmented structures.
A cortical gray matter mask $\M\st{GM}^h : \Omega \to \{0, 1\}$
of hemisphere $h$ is extracted from $\bm{P}$,
where $h$ is randomly chosen from $H = \{\text{left}, \text{right}\}$ with equal probability.

A ``resectable hemisphere mask'' $\M\st{R}^h$ is generated from $\bm{P}$ and $h$ such that $\M\st{R}^h (\p) = 1$ if
${\bm{P}(\p) \neq \{M\st{BG}, M\st{BT}, M\st{CB}, M_{\hat{h}} \} }$
and $0$ otherwise,
where $M\st{BG}$, $M\st{BT}$, $M\st{CB}$ and $M_{\hat{h}}$ are the labels in $Z$ corresponding to the background, brainstem, cerebellum and contralateral hemisphere, respectively.
$\M\st{R}^h$ is smoothed using a series of binary morphological operations, for realism.

A random voxel $\bm{a} \in \Omega$ is selected such that $\M\st{GM}^h(\bm{a}) = 1$.
A translation transform $T\st{T}(\bm{a} - \bm{c})$ is applied to $S\st{E}$ so $S_a = T\st{T}(\bm{a} - \bm{c}) (S\st{E})$ is centered on $\bm{a}$.

A binary image $\binimg{\M_{S_a}}$ is generated from $S_a$ such that $\M_{S_a}(\p) = 1$ for all $\p$ within $S_a$ and $\M_{S_a}(\p) = 0$ outside.
Finally, $\M_{S_a}$ is restricted by $\M\st{R}^h$ to generate the cavity label $\Y\simul = \M_{S_a} \odot \M\st{R}^h$, where $\odot$ represents the Hadamard product.
\cref{fig:shape} illustrates the process.

\subsubsection{Simulating cavities filled with CSF}
\label{sec:texture_cavity}

Brain \acp{rc} are typically filled with \ac{csf}.
To generate a realistic \acs{csf} texture,
we create a ventricle mask
${\M\st{V} : \Omega \to \{ 0, 1 \}}$ from $\bm{P}$, such that
$\M\st{V}(\p) = 1$ for all $\p$ within the ventricles and
$\M\st{V}(\p) = 0$ outside.
Intensity values within the ventricles are assumed to have
a normal distribution~\cite{gudbjartsson_rician_1995}
with a mean $\mu\st{CSF}$ and standard deviation $\sigma\st{CSF}$
calculated from voxel intensity values in
$\{ \X\pre(\p) \mid \p \in \Omega \land \M\st{V}(\p) = 1 \}$.
A \acs{csf}-like image is then generated as $\X\st{CSF}(\p) \sim \NN (\mu\st{CSF}, \sigma\st{CSF}), \forall \p \in \Omega$.

We use $\Y\simul$ to guide blending of $\X\st{CSF}$ and $\X\pre$ as follows.
A Gaussian filter is applied to $\Y\simul$ to obtain a smooth alpha channel $\img{\AAA_\alpha}{[0, 1]}$ defined as
$
    \AAA_\alpha
    = \Y\simul
    * \bm{G}_{\NN}(\bm{\sigma}),
$
where
$*$ is the convolution operator
and $\bm{G}_{\NN}(\bm{\sigma})$ is a 3D Gaussian kernel with standard deviations
$\bm{\sigma} = (\sigma_x, \sigma_y, \sigma_z)$.
Then, $\X\st{CSF}$ and $\X\pre$ are blended by the convex combination
\begin{equation}
    \X\simul
    = \AAA_\alpha \odot \X\st{CSF}
    + (1 - \AAA_\alpha) \odot \X\pre
\end{equation}

\begin{figure}
    \centering
    \captionsetup[subfigure]{aboveskip=3pt, belowskip=5pt}

    \begin{subfigure}{0.16\textwidth}
        \includegraphics[width=0.99\linewidth]{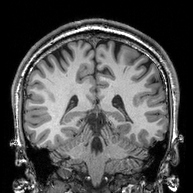}
        \caption{\label{fig:tmri}}
    \end{subfigure}
    \begin{subfigure}{0.16\textwidth}
        \includegraphics[width=0.99\linewidth]{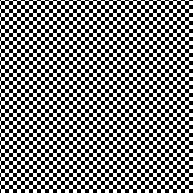}
        \caption{\label{fig:checkerboard}}
    \end{subfigure}
    \begin{subfigure}{0.16\textwidth}
        \includegraphics[width=0.99\linewidth]{Mr}
        \caption{\label{fig:tmh}}
    \end{subfigure}
    \begin{subfigure}{0.16\textwidth}
        \includegraphics[width=0.99\linewidth]{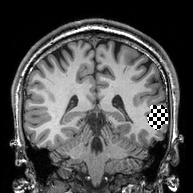}
        \caption{\label{fig:blh}}
    \end{subfigure}
    \begin{subfigure}{0.16\textwidth}
        \includegraphics[width=0.99\linewidth]{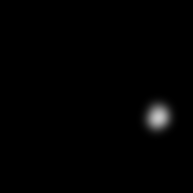}
        \caption{\label{fig:tms}}
    \end{subfigure}
    \begin{subfigure}{0.16\textwidth}
        \includegraphics[width=0.99\linewidth]{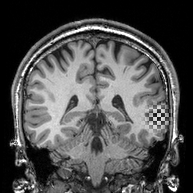}
        \caption{\label{fig:bls}}
    \end{subfigure}

    \caption{%
        Simulation of resected image $\X\simul$.
        We use a checkerboard for visualization.
        Two scalar-valued images $\X\pre$ (\subref{fig:tmri})
        and $\X_2$ (\subref{fig:checkerboard})
        are blended using $\Y\simul$ (\subref{fig:tmh})
        and $\sigma_i = \SI{0}{\milli \meter}$ to create an image with hard boundaries (\subref{fig:blh})
        and $\sigma_i = \SI{5}{\milli \meter}$ (\subref{fig:tms})
        for an image with soft boundaries (\subref{fig:bls}),
        mimicking partial-volume effects
    }
    \label{fig:texture}
\end{figure}

We use $\bm{\sigma} > 0$ to mimic partial-volume effects at the cavity boundary.
The blending process is illustrated in \cref{fig:texture}.

\section{Experiments and results}
\label{sec:experiments_and_results}

\subsection{Data}
\label{sec:data}
\subsubsection{Public data for simulation}

\ac{t1} \acp{mri} were collected from publicly available datasets \ac{ixi}, \ac{adni}, and \ac{oasis}, for a total of 1813 images.
They are used as control subjects in our self-supervised experiments (\cref{sec:sim_res_self}).
Note that, although we use the term ``control'' to refer to subjects that have not undergone resective surgery, they may have other neurological conditions.
For example, subjects in \ac{adni} may suffer from \ac{ad}.

\subsubsection{Multicenter epilepsy data}
\label{sec:multicenter}

We evaluate the generalizability of our approach to data from several institutions: \textit{Milan} ($n = 20$), \textit{Paris} ($n = 19$), \textit{Strasbourg} ($n = 33$), and EPISURG ($n = 133$).
We curated the EPISURG dataset from patients with refractory focal epilepsy who underwent resective surgery between 1990 and 2018 at the \ac{nhnn}, London, United Kingdom.
All images in EPISURG were defaced using a predefined face mask in the \ac{mni} space to preserve patient identity.
In total, there were 430 patients with postoperative \ac{t1} \ac{mri}, 268 of which had a corresponding preoperative \ac{mri}.
EPISURG is available online and can be freely downloaded~\cite{perez-garcia_episurg_2020}.
The same human rater (F.P.G.) annotated all images
semi-automatically using 3D Slicer 4.10~\cite{fedorov_3d_2012}.

\subsubsection{Brain tumor datasets}

The \ac{bite} dataset~\cite{mercier_online_2012} consists of ultrasound and \ac{mri} of patients with brain tumors.
We use 13 postoperative \ac{t1gad} to perform a qualitative assessment of our model's generalization to images from a substantially different domain (contrast-enhanced images) and different pathology, where different surgical techniques may affect \ac{rc} appearance.

\subsubsection{Preprocessing}
\label{sec:preprocessing}

\newcommand{\zoom}[3]{$ #1 \times #2 \times #3 $}

For all images, the brain was segmented using ROBEX~\cite{iglesias_robust_2011}.
They were resampled into the \ac{mni} space using sinc interpolation to preserve image quality.
After resampling, all images had a 1-mm isotropic resolution and size \zoom{193}{229}{193}.

\subsection{Network architecture and implementation details}
We used the PyTorch deep learning framework, training with \ac{amp} on two 32-GB TESLA V100 GPUs.
We used Sacred~\cite{greff_sacred_2017} to configure, log and visualize experiments.

We implemented a 3D U-Net~\cite{cicek_3d_2016} variant using two contractive and expansive blocks, upsampling with trilinear interpolation for the synthesis path, and 1/4 of the filters for each convolutional layer.
We used dilated convolutions, starting with a dilation factor of one, then increased or decreased in steps of one after each contractive or expansive block, respectively.
Our architecture has the same receptive field ($\SI{88}{\milli \meter}^3$) but $\approx 77 \times$ fewer parameters (\num{246156}) than the original 3D U-Net, reducing overfitting and computational burden.

Convolutional layers were initialized using He's method, and followed by batch normalization and nonlinear PReLU activation functions.
We used adaptive moment estimation (AdamW) to adjust the learning rate, with
weight decay of~$10^{-2}$,
and a learning scheduler that divides the learning rate by ten every 20 epochs.
We optimized our network to minimize the mean soft Dice loss of each mini-batch.
For training, a mini-batch size of ten images (five per GPU) was used.
Self-supervised training took approximately 27 hours.
Fine-tuning on a small annotated dataset took approximately seven hours.

\subsection{Processing during training}
\label{sec:preprocessing_augmentation}
We use TorchIO transforms to load, preprocess and augment our data during training~\cite{perez-garcia_torchio_2020}.
Instead of preprocessing images with denoising or bias removal, we simulate different artifacts in the training instances so that our models are robust to artifacts.
Our preprocessing and augmentation transforms are:
1) \ac{rs} of resections (self-supervised training only),
2) histogram standardization,
3) Gaussian blurring or \ac{rs} of anisotropic spacing,
4) \ac{rs} of \ac{mri} ghosting,
5) spike and
6) motion artifacts,
7) \ac{rs} of bias field inhomogeneity,
8) standardization of foreground to zero-mean and unit variance,
9) Gaussian noise,
10) \ac{rs} of affine or free-form transformations,
11) random flip around the sagittal plane,
and 12) crop to a tight bounding box around the brain.
We refer the reader to our GitHub repository for details.

\subsection{Experiments}

Overlap measurements are reported as `median (interquartile range)' \ac{dsc}.
No postprocessing is performed for evaluation, except thresholding at 0.5.
We analyzed differences in model performance using a one-tailed Mann-Whitney $U$ test (as \acp{dsc} were not normally distributed) with a significance threshold of $\alpha = 0.05$, and Bonferroni correction for $n$ experiments: $\alpha\st{Bonf} = \frac{\alpha}{n (n - 1)}$.

\subsubsection{Self-supervised learning: training with simulated resections only}
\label{sec:self}

In our first experiment, we assess the relation between the resection simulation complexity and the segmentation performance of the model.
We train our model with simulated resections on the publicly available dataset
$D\pre = \{ \X_{\text{preop}_i} \}_{i = 1}^{n\pre}$, where $n\pre = 1813$ (\cref{sec:data}).
We use 90\% of the images in $D\pre$ for the training set $D\st{pre,train}$ and 10\% for the validation set.
At each training iteration, $b$ images from $D\st{pre,train}$ are loaded, resected, preprocessed and augmented to obtain a mini-batch of $b$ training instances
$\{ ( \X_{\text{sim}_i}, \Y_{\text{sim}_i}) \}_{i = 1}^{b}$.
Note that the resection simulation is performed on the fly, which ensures that the network never sees the same resection during training.
Models were trained for 60 epochs, using an initial learning rate of $10^{-3}$.
We use the model weights from the epoch with the lowest mean validation loss obtained during training for evaluation.
Models were tested on the 133 annotated images in EPISURG.

\begin{figure}
    \centering
    \captionsetup[subfigure]{aboveskip=3pt, belowskip=5pt}

    \begin{subfigure}{0.26\textwidth}
        \includegraphics[width=\linewidth]{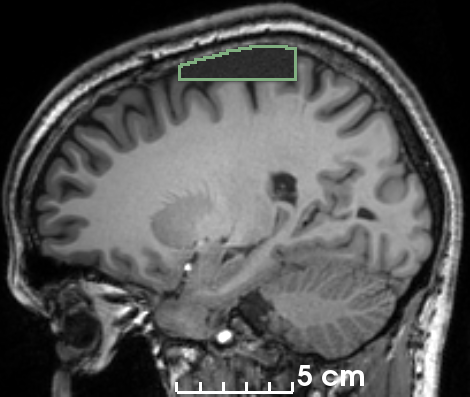}
        \caption{\label{fig:exp_shape_cuboid}}
    \end{subfigure}
    \hfill
    \begin{subfigure}{0.26\textwidth}
        \includegraphics[width=\linewidth]{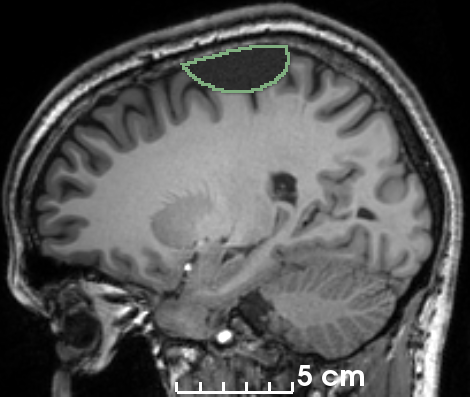}
        \caption{\label{fig:exp_shape_ellipsoid}}
    \end{subfigure}
    \hfill
    \begin{subfigure}{0.26\textwidth}
        \includegraphics[width=\linewidth]{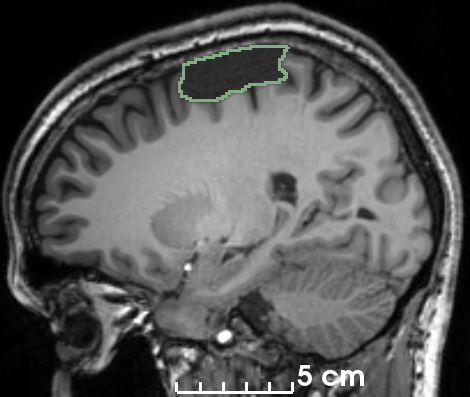}
        \caption{\label{fig:exp_shape_noisy}}
    \end{subfigure}

    \caption{%
        Simulation of \acp{rc} with increasing shape complexity (\cref{sec:simulation}):
        cuboid (\subref{fig:exp_shape_cuboid}),
        ellipsoid (\subref{fig:exp_shape_ellipsoid})
        and ellipsoid perturbed with simplex noise (\subref{fig:exp_shape_noisy})
    }
    \label{fig:exp_shape}
\end{figure}

To investigate the effect of the simulated cavity shape on model performance, we modify $\phi\simul$ to generate cuboid- (\cref{fig:exp_shape_cuboid}) or ellipsoid-shaped (\cref{fig:exp_shape_ellipsoid}) resections, and compare with the baseline ``noisy'' ellipsoid (\cref{fig:exp_shape_noisy}).
The cuboids and ellipsoid meshes are not perturbed using simplex noise, and cuboids are not rotated.

Best results were obtained by the baseline model (80.5 (18.7)), trained using ellipsoids perturbed with procedural noise.
Models trained with cuboids and rotated ellipsoids performed significantly (57.9 (73.1), $p < 10^{-8}$) and marginally (79.0 (20.0), $p = 0.123$) worse.

% $ seed=21 && shape="noisy" && debug_dir="/tmp/debug_dir_${seed}_${shape}" && resect /Users/fernando/git/ijcars-2020/nii/t1.nii.gz /Users/fernando/git/ijcars-2020/nii/t1_seg_gif.nii.gz $debug_dir/resected.nii.gz $debug_dir/label.nii.gz -miv 20000 -mav 20000 -d $debug_dir -s $seed -r 25 -18 74 --shape $shape

% Slice 119

\subsubsection{Fine-tuning on small clinical datasets}

We assessed the generalizability of our baseline model by fine-tuning it on small datasets from four institutions that may use different surgical approaches and acquisition protocols (including contrast enhancement and anisotropic spacing in \textit{Strasbourg}) (\cref{sec:multicenter}).
Additionally, we fine-tuned the model on 20 cases from EPISURG with the lowest \ac{dsc} in \cref{sec:self}.

For each dataset, we load the pretrained baseline model, initialize the optimizer with an initial learning rate of $5 \times 10^{-4}$, initialize the learning rate scheduler and fine-tune all layers simultaneously for 40 epochs using 5-fold cross-validation.
We use model weights from the epoch with the lowest mean validation loss for evaluation.
To minimize data leakage, we determined the above hyperparameters using the validation set of one fold in the \textit{Milan} dataset.

\begin{figure}
    \centering
    \includegraphics[width=\linewidth]{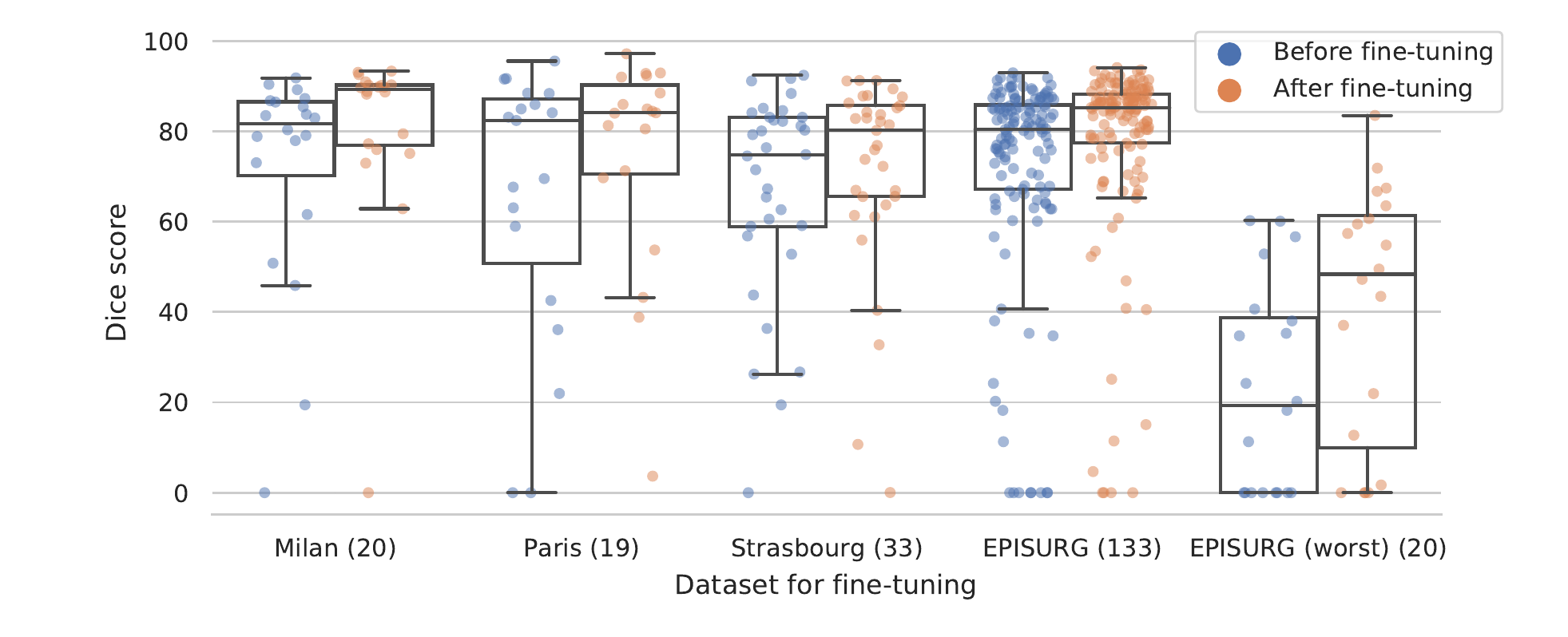}
    \caption{%
        \ac{dsc} without (blue) and with (orange) fine-tuning of the model training using self-supervision.
        Horizontal lines in the boxes represent the first, second (median) and third quartiles.
        \textcolor{rev1}{\textit{EPISURG (worst)} comprises the 20 cases from EPISURG with the lowest \ac{dsc} in the experiment described in \cref{sec:self}.}
        Numbers in parentheses indicate subjects per dataset
    }
    \label{fig:finetuning_quant}
\end{figure}

We observed a consistent increase in \ac{dsc} for all fine-tuned models, up to a maximum of 89.2 (13.3) for the \textit{Milan} dataset.
For comparison, inter-rater agreement between human annotators in our previous study was 84.0 (9.9)~\cite{perez-garcia_simulation_2020}.
Quantitative evaluation is illustrated in \cref{fig:finetuning_quant}.

\subsubsection{Qualitative evaluation on brain tumor resection dataset}

% Failures: 3 and X
% Mention the failures?

We used the \ac{bite} dataset~\cite{mercier_online_2012} to evaluate the ability of our self-supervised model to segment \acp{rc} on images from a different institution, modality and pathology than the datasets used for quantitative evaluation.
For postprocessing, all but the largest binary connected component were removed.
The model successfully segmented the \ac{rc} on 11/13 images, even though some contained challenging features (\cref{fig:bite}).

\newcommand{\qualit}[1]{\includegraphics[width=0.49\linewidth]{bite/cropped/#1}}
% https://tex.stackexchange.com/a/381477/216202
\begin{figure}
    \centering
    \begin{subfigure}{0.32\textwidth}
        \centering
        \qualit{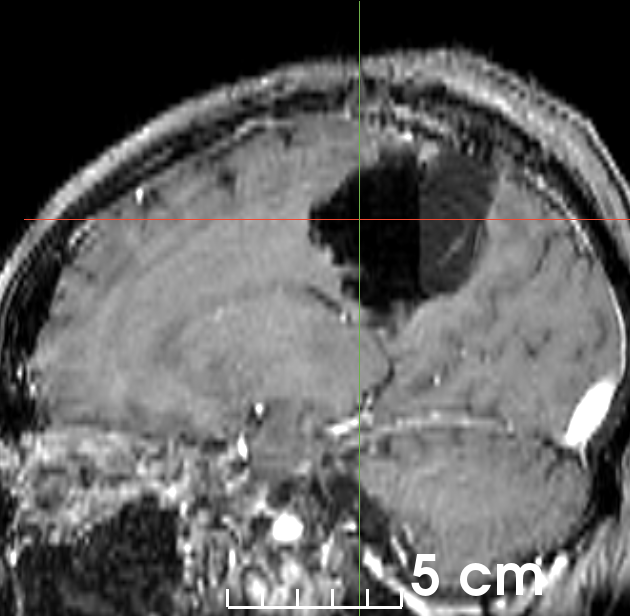}%
        \hfill
        \qualit{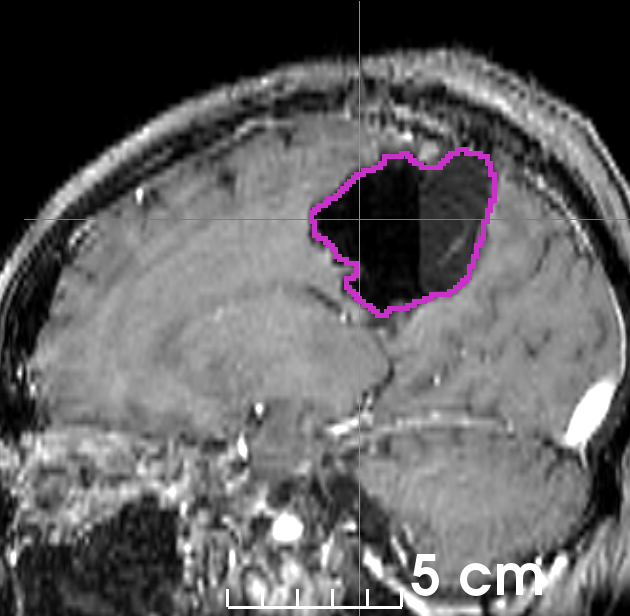}
        \caption{\label{fig:bite_air}}
    \end{subfigure}%
    \hfill
    \begin{subfigure}{0.32\textwidth}
        \centering
        \qualit{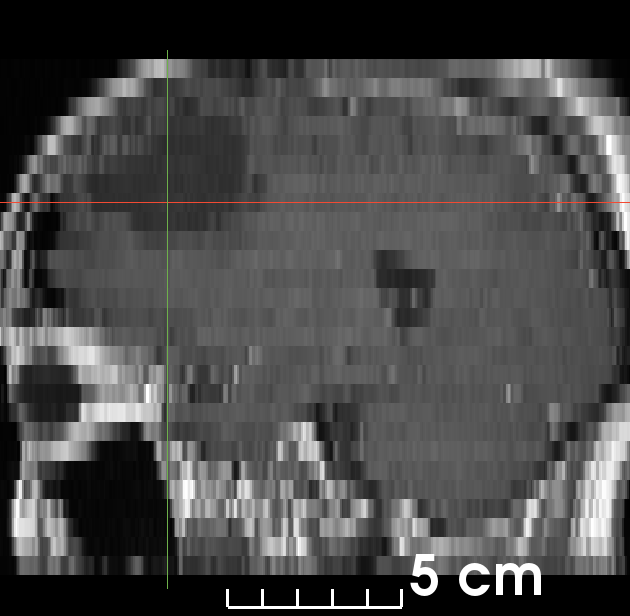}%
        \hfill
        \qualit{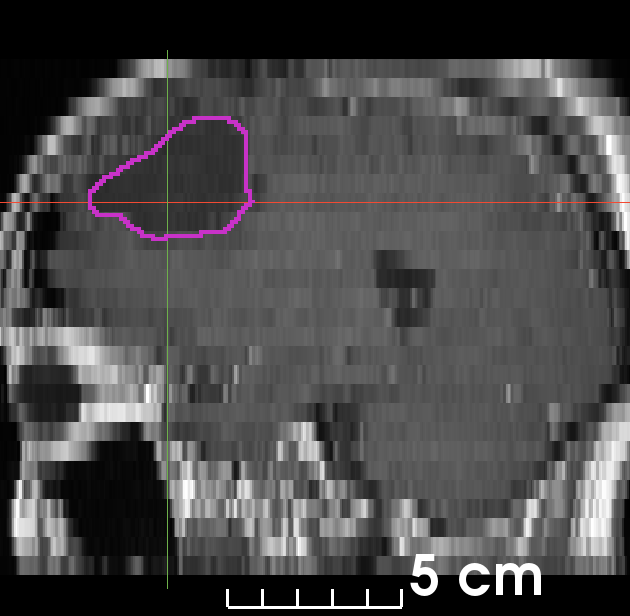}
        \caption{\label{fig:bite_aniso}}
    \end{subfigure}%
    \hfill
    \begin{subfigure}{0.32\textwidth}
        \centering
        \qualit{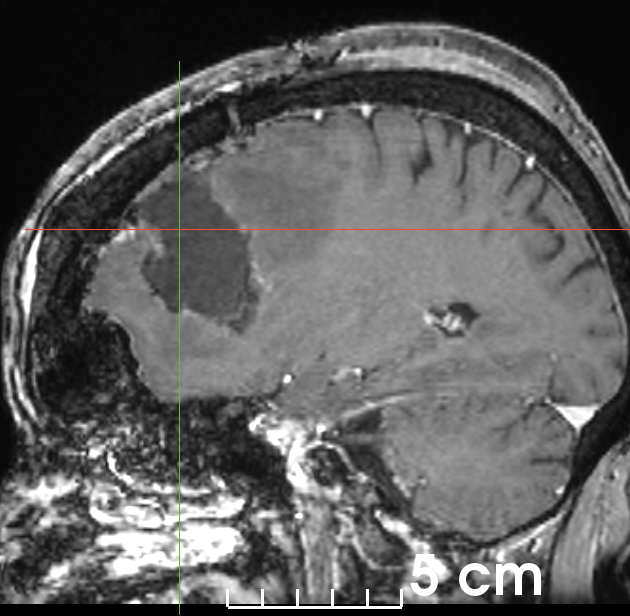}%
        \hfill
        \qualit{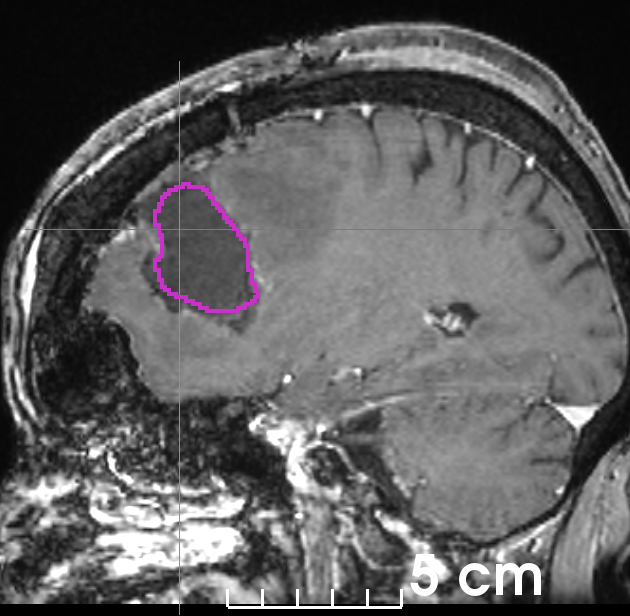}
        \caption{\label{fig:bite_motion}}
    \end{subfigure}
    \caption{%
        Qualitative results on postoperative brain tumor \ac{t1gad} \ac{mri}.
        The model is robust to:
        air and \ac{csf} in the \ac{rc} (\subref{fig:bite_air}),
        anisotropic spacing (\subref{fig:bite_aniso}),
        presence of edema (\subref{fig:bite_motion}),
        and a different modality than used for training (all).
        \textcolor{rev1}{Note that these images are from a different institution, modality and pathology than the datasets used for quantitative evaluation.
        Manual annotations are not available}
    }
    \label{fig:bite}
\end{figure}

\subsubsection{Qualitative evaluation on intraoperative image}

\begin{figure}
    \centering
    \includegraphics[trim=0 0 0 12, clip, width=0.8\linewidth]{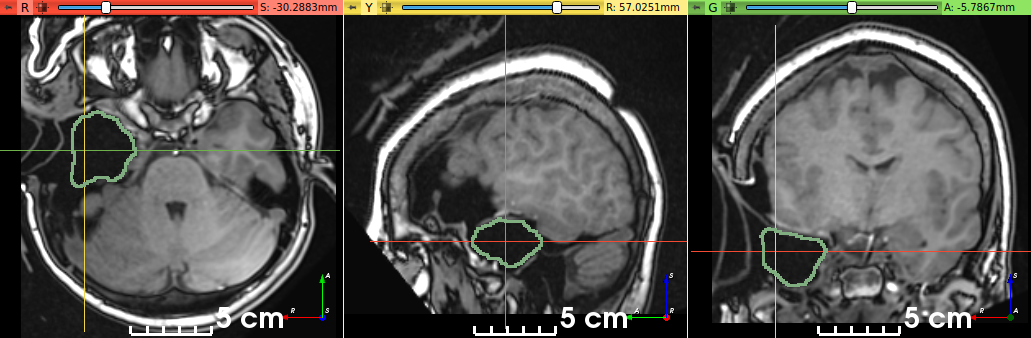}
    \caption{%
        Qualitative result on an intraoperative \ac{mri}.
        The baseline model correctly discarded regions filled with air or \ac{csf} outside of the \ac{rc}
    }
    \label{fig:intra}
\end{figure}

We used our baseline model to segment the \ac{rc} on one intraoperative \ac{mri} from our institution.
Despite the large domain shift between the training dataset and the intraoperative image, which includes a retracted skin flap and a missing bone flap, the model was able to correctly estimate the \ac{rc}, discarding similar regions filled with \ac{csf} or air (\cref{fig:intra}).

\section{Discussion and conclusion}
\label{sec:discussion}

We addressed the challenge of segmenting postoperative brain resection cavities from \ac{t1} \ac{mri} without annotated data.
We developed a self-supervised learning strategy to train without manually annotated data, and a method to simulate \acp{rc} from preoperative \ac{mri} to generate training data.
Our novel approach is conceptually simple, easy to implement, and relies on clinical knowledge about postoperative phenomena.
The resection simulation is computationally efficient ($< \SI{1}{\second}$), so it can run during training as part of a data augmentation pipeline.
It is implemented within the TorchIO framework~\cite{perez-garcia_torchio_2020} to leverage other data argumentation techniques during training, enabling our model to have a robust performance across \ac{mri} of variable quality.

Modeling a realistic cavity shape is important (\cref{sec:self}).
Our model generalizes well to clinical data from different institutions and pathologies, including epilepsy and glioma.
Models may be easily fine-tuned using small annotated clinical datasets to improve performance.
Moreover, our resection simulation and learning strategy may be extended to train with arbitrary modalities, or synthetic modalities generated from brain parcellations~\cite{billot_learning_2020}.
Therefore, our strategy can be adopted by institutions with a large amount of unlabeled data, while fine-tuning and testing on a smaller labeled dataset.

Poor segmentation performance is often due to very small cavities, where the cavity was not detected, and large brain shift or subdural edema, where regions were incorrectly segmented.
The former issue may be overcome by training with a distribution of cavity volumes which oversamples small resections.
The latter can be addressed by extending our method to simulate displacement with biomechanical models or nonlinear deformations of the brain~\cite{granados_generative_2021}.

We showed that our model correctly segmented an intraoperative image, respecting imaginary boundaries between brain and skull, suggesting a good inductive bias of human neuroanatomy.
Qualitative results and execution time, which is in the order of milliseconds, suggest that our method could be used intraoperatively\textcolor{rev1}{, for image guidance during resection or} to improve registration with preoperative images by masking the cost function using the \ac{rc} segmentation~\cite{brett_spatial_2001}.
\textcolor{rev1}{Segmenting the \ac{rc} may also be used to study potential damage to white matter tracts postoperatively~\cite{winston_optic_2012}.
Our method could be easily adapted to simulate other lesions for self-supervised training, such as cerebral microbleeds~\cite{cuadrado-godia_cerebral_2018}, narrow and snake-shaped \acp{rc} typical of disconnective surgeries~\cite{mohamed_temporoparietooccipital_2011}, or \acp{rc} with residual tumor~\cite{meier_automatic_2017}.
}

As part of this work, we curated and released EPISURG, an \ac{mri} dataset with annotations from three independent raters.
EPISURG could serve as a benchmark dataset for quantitative analysis of pre- and postoperative imaging of open resection for epilepsy treatment.
To the best of our knowledge, this is the first open annotated database of post-resection \ac{mri} for epilepsy patients.

    % https://www.springer.com/journal/11548/submission-guidelines#Instructions%20for%20Authors_Text
% Acknowledgments
% Acknowledgments of people, grants, funds, etc. should be placed in a separate section on the title page. The names of funding organizations should be written in full.

\begin{acknowledgements}
    Some of the data used in preparation of this article was obtained from the Alzheimer’s Disease Neuroimaging Initiative (ADNI) database.
    As such, the investigators within the ADNI contributed to the design and implementation of ADNI and/or provided data but did not participate in analysis or writing of this report.
    A complete listing of ADNI investigators can be found at \myhref{http://adni.loni.usc.edu/}.
    \ac{ixi} can be found at \myhref{https://brain-development.org/ixi-dataset/}.
    Data were also provided in part by the Open Access Series of Imaging Studies (OASIS) (\myhref{https://www.oasis-brains.org/}).
    We thank Philip Noonan and David Drobny for the fruitful discussions and feedback.
    \textbf{Availability of data and material}
EPISURG can be freely downloaded from the UCL Research Data Repository~\cite{perez-garcia_episurg_2020}.
\textbf{Code availability}
The code for resection simulation, training and inference is available at \myhref{https://github.com/fepegar/resseg-ijcars}.
A tool to segment \acp{rc} using our best model (\cref{sec:self}) can be installed from the \ac{pypi}: \texttt{pip install resseg}.
\textbf{Authors' contributions}
\textit{Conceptualization}: F.P.G., R.S., J.S.D. and S.O.;
\textit{Methodology}: F.P.G. and R.S.;
\textit{Software, Formal Analysis, Investigation} and \textit{Visualization}: F.P.G.;
\textit{Resources}: F.P.G., M.R., F.C., V.F., V.N., C.E., I.O. and J.S.D.;
\textit{Data Curation}: F.P.G. and J.S.D.;
\textit{Writing --- Original Draft}: F.P.G.;
\textit{Writing -- Review \& Editing}: F.P.G., R.D., R.S., J.S.D. and S.O.;
\textit{Supervision}: T.V., R.S., J.S.D. and S.O.;
\textit{Project Administration}: J.S.D. and S.O.;
\textit{Funding Acquisition}: R.S., J.S.D. and S.O.
\textbf{Funding}
This publication represents, in part, independent research commissioned by the Wellcome Innovator Award (218380/Z/19/Z/).
Computing infrastructure at the Wellcome / EPSRC Centre for Interventional and Surgical Sciences (WEISS) (UCL) (203145Z/16/Z) was used for this study.
R.D. is supported by the Wellcome Trust (203148/Z/16/Z) and the Engineering and Physical Sciences Research Council (EPSRC) (NS/A000049/1).
T.V. is supported by a Medtronic / Royal Academy of Engineering Research Chair (RCSRF1819\textbackslash7\textbackslash34).
The views expressed in this publication are those of the authors and not necessarily those of the Wellcome Trust.
\textbf{Conflicts of interest}
The authors declare that they have no conflict of interest.
\textbf{Research involving human participants}
All procedures performed in studies involving human participants were in accordance with the ethical standards of the institutional and/or national research committee and with the 1964 Helsinki declaration and its later amendments or comparable ethical standards.
\textbf{Informed Consent}
For this type of study, formal consent was not required.

\end{acknowledgements}

    \bibliographystyle{spmpsci}      % mathematics and physical sciences

% Guidelines say: Authors preparing their manuscript in LaTeX can use the bibtex file spbasic.bst which is included in Springer’s LaTeX macro package.
% But this lists all authors at each citation!
%\bibliographystyle{spbasic}

\bibliography{MICCAI_2020}   % name your BibTeX data base

\end{document}